\documentclass[pra,onecolumn,showpacs]{revtex4}
\usepackage{graphicx}
\usepackage{dcolumn}
\usepackage{bm}
\usepackage{color}
\usepackage{amsmath}
\usepackage{amsfonts}
\usepackage{amssymb}

\input{tcilatex}
\begin{document}

\title{Quantum mechanics of inverted potential well \\
--- Hermitian Hamiltonian with imaginary eigenvalues, quantum-classical
correspondence}
\author{Ni Liu}
\email{liuni2011520@sxu.edu.cn}
\affiliation{Institute of Theoretical Physics, State Key Laboratory of Quantum Optics and
Quantum Optics Devices, Shanxi University, Taiyuan 030006, Shanxi, China}
\author{J.-Q. Liang}
\email{jqliang@sxu.edu.cn} \affiliation{Institute of Theoretical
Physics, State Key Laboratory of Quantum Optics and Quantum Optics
Devices, Shanxi University, Taiyuan 030006, Shanxi, China}

\begin{abstract}
We in this paper study the quantization of a particle in an inverted
potential well. The Hamiltonian is Hermitian, while the potential is
unbounded below. Classically the particle moves away acceleratingly from the
center of potential top. The existing eigenstates must be unstable with
imaginary eigenvalues, which characterize the decay rate of states. We solve
the Hamiltonian problem of inverted potential well by the algebraic method
with imaginary-frequency raising and lowering boson operators similar to the
normal oscillator case. The boson number operator is non-Hermitian, while
the integer-number eigenvalues are, of course, real. Dual sets of
eigenstates, denoted by "bra" and "ket", are requested corresponding
respectively to the complex conjugate number-operators. Orthonormal
condition exists between the "bra" and "ket" states. We derive a spatially
non-localized generating function, from which $n$-th eigenfunctions can be
generated by the raising operators in coordinate representation. The "bra"
and "ket" generating functions are mutually normalized with the imaginary
integration measure. The probability density operators defined between the
"bra" and "ket" states are non-Hermitian invariants, which lead to the Schr%
\H{o}dinger equations respectively for the "bra" and "ket" states. While
probabilities of "bra" and "ket" states themselves are not conserved
quantities because of the decay. The imaginary-frequency boson coherent
states are defined as eigenstates of lowering operators. The minimum
uncertainty relation is proved explicitly in the coherent states. Finally
the probability average of Heisenberg equation in the coherent states is
shown precisely in agreement with the classical equation of motion. The
quantum-classical correspondence exists in the imaginary eigenvalue system.
\end{abstract}

\maketitle
\date{\today }
\date{\today }
\date{\today }
\date{\today }
\date{\today }
\date{\today }
\date{\today }
\date{\today }

\section{Introduction}

It is a common knowledge that the Hamiltonian in quantum mechanics is
Hermitian to guarantee real energy spectrum and the conservation of
probability. The non-Hermitian Hamiltonian describes usually open systems
with complex energy spectrum. Bender and his collaborators discovered for
the first time that the Hermitian Hamiltonian is sufficient but not
necessary to have real eigenvalues \cite{CM98,CM99,CM02}. The non-Hermitian
Hamiltonian possesses real spectrum if it has the parity-time ($PT$)
reflection symmetry \cite%
{CM98,CM99,CM02,CM13,CM04,CM07,CM18,GuResult,GuAnn,LiuResult,DN18}. The
non-Hermitian Hamiltonian with real spectrum is called pseudo-Hermitian by
Mostafazadeh, who derived a necessary condition \cite{Mosta1,Mosta2,Mosta3}
of it. The pseudo-Hermitian Hamiltonian, however, is not confined by the $PT$%
-symmetry. It is found that the $PT$ -asymmetric Hamiltonian can also
possess real spectra \cite{GuResult,LiuPysica,Amaa,PRA}.

As a matter of fact not only the non-Hermitian Hamiltonians but also the
Hermitian Hamiltonian of a conservative system can have complex spectrum if
the potential is unbounded below. The complex spectrum of an inverted
double-well potential was studied long ago in terms of the instanton method
under semiclassical approximation \cite{LM92,LM94}. The probability density
of eigenstates is no longer a conserved quantity different from the real
eigenvalues. The imaginary part of energy eigenvalues characterizes the
decay rate or life time of metastable states \cite{LM92,LM94}. As of yet the
rigorous complex eigenvalues and associated eigenstates have not been given
in the framework of quantum mechanics for the Hermitian Hamiltonians.

We solve in the present paper the Hamiltonian problem for a particle in an
inverted potential well by means of algebraic method.

Classically the particle moves acceleratingly away from the center of the
inverted potential well. The quantum wave functions are, of course,
spatially non-localized.

\section{Eigenstates for a particle in the inverted potential well}

We consider the Hamiltonian for a particle of unit mass $m=1$ in an inverted
potential well%
\begin{equation}
H=\frac{p^{2}}{2}-\frac{1}{2}\omega ^{2}x^{2}.  \label{c}
\end{equation}%
The solution of classical equation is%
\begin{equation}
x\left( t\right) =\pm \frac{v}{\omega }\sinh \omega t,  \label{c-1}
\end{equation}%
where $v$ is the initial velocity. The particle moves acceleratingly away
from the center of inverted potential well. Without the initial velocity ($%
v=0$) the particle would stay at the top of potential forever. While quantum
mechanically it will fall down because of the uncertainty principle. This
unstable static-solution has been well studied from viewpoint of quantum
mechanics \cite{Mull}. It is known as spheleron in $\phi ^{4}$ field theory
\cite{LW2023}. The existing eigenstates must be unstable with pure imaginary
eigenvalues.

\subsection{Complex eigenvalues of Hermitian Hamiltonian}

First of all we present the general formalism of quantum mechanics for the
Hermitian Hamiltonian with complex eigenvalues. Dual sets of eigenstates are
required respectively for the discrete conjugate eigenvalues such
that,\bigskip
\begin{equation}
\widehat{H}|u_{n}\rangle _{r}=E_{n}|u_{n}\rangle _{r},\quad \widehat{H}%
|u_{n}\rangle _{l}=E_{n}^{\ast }|u_{n}\rangle _{l},  \label{ei}
\end{equation}%
where $|u_{n}\rangle _{i}$ with $i=r,l$ are called respectively the "ket"
and "bra" states. It is called the normal order that Hamiltonian acts on the
right-hand eigenstates in Eq.(\ref{ei}). Taking complex conjugate of the Eq.(%
\ref{ei}) we have the anti-normal order action%
\begin{equation}
_{r}\langle u_{n}|\widehat{H}=\quad _{r}\langle u_{n}|E_{n}^{\ast },\quad
_{l}\langle u_{n}|\widehat{H}=\quad _{l}\langle u_{n}|E_{n}.  \label{ei-1}
\end{equation}%
The orthonormal condition is%
\begin{equation}
_{l}\langle u_{n}|u_{m}\rangle _{r}=\delta _{nm}  \label{n}
\end{equation}%
between dual-set eigenstates. The expectation values
\begin{equation*}
_{l}\langle u_{n}|\widehat{H|}u_{n}\rangle _{r}=E_{n},\quad _{r}\langle
u_{n}|\widehat{H|}u_{n}\rangle _{l}=E_{n}^{\ast },
\end{equation*}%
are well defined. While the expectation values%
\begin{eqnarray*}
_{r}\langle u_{n}|\overrightarrow{H}|u_{n}\rangle _{r} &=&E_{n}\quad
_{r}\langle u_{n}|u_{n}\rangle _{r},\quad _{r}\langle u_{n}|\overleftarrow{H}%
|u_{n}\rangle _{r}=E_{n}^{\ast }\quad _{r}\langle u_{n}|u_{n}\rangle _{r}, \\
_{l}\langle u_{n}|\overrightarrow{H}|u_{n}\rangle _{l} &=&E_{n}^{\ast }\quad
_{l}\langle u_{n}|u_{n}\rangle _{l},\quad _{l}\langle u_{n}|\overleftarrow{H|%
}u_{n}\rangle _{l}=E_{n}\quad _{l}\langle u_{n}|u_{n}\rangle _{l},
\end{eqnarray*}%
depend on the normal-order Eq.(\ref{ei}) and anti-normal-order Eq.(\ref{ei-1}%
) actions. The state density $_{i}\langle u_{n}|u_{n}\rangle _{i}$ for both
"bra" and "ket" states $i=r,l$ is non-normalizable..

The special solutions of Schr$\ddot{o}$dinger equations

\begin{equation}
i\hbar \frac{\partial }{\partial t}|\psi \rangle _{i}=\widehat{H}|\psi
\rangle _{i}  \label{s}
\end{equation}%
for $i=r,l$ are obviously%
\begin{eqnarray}
|\psi _{n}\left( t\right) \rangle _{r} &=&e^{\frac{1}{\hbar }\left(
-iE_{n}^{R}+E_{n}^{I}\right) t}|u_{n}\rangle _{r}  \notag \\
|\psi _{n}\left( t\right) \rangle _{l} &=&e^{\frac{1}{\hbar }\left(
-iE_{n}^{R}-E_{n}^{I}\right) t}|u_{n}\rangle _{l}  \label{s-1}
\end{eqnarray}%
where the complex energy eigenvalue is written as real and imaginary parts $%
E_{n}=E_{n}^{R}+iE_{n}^{I}$. The probability densities of the "ket" and
"bra" states
\begin{eqnarray*}
_{r}\langle \psi _{n}\left( t\right) |\psi _{n}\left( t\right) \rangle _{r}
&=&e^{\frac{2}{\hbar }E_{n}^{I}t}{}_{r}\langle u_{n}|u_{n}\rangle _{r}, \\
_{l}\langle \psi _{n}\left( t\right) |\psi _{n}\left( t\right) \rangle _{l}
&=&e^{-\frac{2}{\hbar }E_{n}^{I}t}{}_{l}\langle u_{n}|u_{n}\rangle _{l}
\end{eqnarray*}%
are not the conserved quantities either. While the orthonormal condition Eq.(%
\ref{n}) is valid for the time-dependent states%
\begin{equation*}
_{l}\langle \psi _{m}\left( t\right) |\psi _{n}\left( t\right) \rangle
_{r}=\delta _{mn}.
\end{equation*}

\subsection{Non-Hermitian number operator and imaginary eigenvalues of
Hermitian Hamiltonian}

For the inverted potential well any eigenstate is unstable and thus
eigenvalue, if it exists, must be pure imaginary. We are going to solve the
potential model by means of algebraic method. The Hamiltonian can be
represented as%
\begin{equation}
\widehat{H}=i\omega \hbar \left[ \frac{\widehat{p}^{2}}{2i}+\frac{1}{2}i%
\widehat{x}^{2}\right] ,  \label{A}
\end{equation}%
\textbf{\ }in which\textbf{\ }$\widehat{x}$, $\widehat{p}$ are dimensionless
variables. The imaginary frequency boson operators defined by
\begin{equation}
\widehat{a}_{-}=\sqrt{\frac{i}{2}}\left( \widehat{x}+\widehat{p}\right)
,\quad \widehat{a}_{+}=\sqrt{\frac{i}{2}}\left( \widehat{x}-\widehat{p}%
\right)  \label{A-2}
\end{equation}%
satisfy the usual commutation relation%
\begin{equation}
\left[ \widehat{a}_{-},\widehat{a}_{+}\right] =1,  \label{A-1}
\end{equation}%
While
\begin{equation*}
\widehat{a}_{-}^{\dag }=i\widehat{a}_{-},\quad \widehat{a}_{+}^{\dag }=i%
\widehat{a}_{+}.
\end{equation*}%
The Hamiltonian Eq.(\ref{A}) becomes in terms of the boson operators
\begin{equation}
\widehat{H}=i\omega \hbar \left( \widehat{n}+\frac{1}{2}\right) ,  \label{B}
\end{equation}%
in which the boson number operator%
\begin{equation*}
\widehat{n}=\widehat{a}_{+}\widehat{a}_{-}
\end{equation*}%
is non-Hermitian since%
\begin{equation}
\widehat{n}^{\dag }=-\widehat{a}_{-}\widehat{a}_{+}=-\left( \widehat{n}%
+1\right) .  \label{C}
\end{equation}%
While the Hamiltonian Eq.(\ref{B}) is Hermitian%
\begin{equation*}
\widehat{H}^{\dag }=\widehat{H}
\end{equation*}%
The non-Hermitian number operator with real eigenvalues is called the
pseudo-Hermitian \cite%
{Mosta1,Mosta2,Mosta3,GuResult,LiuPysica,Amaa,LM92,LM94,PRA}, which
possesses also the dual-set of eigenstates%
\begin{eqnarray}
\widehat{n}|n\rangle _{r} &=&n|n\rangle _{r};\quad _{r}\langle n|n=\quad
_{r}\langle n|\widehat{n}^{\dag }  \notag \\
\widehat{n}^{\dag }|n\rangle _{l} &=&n|n\rangle _{l};\quad _{l}\langle
n|n=\quad _{l}\langle n|\widehat{n}  \label{C-2}
\end{eqnarray}%
The orthonormal condition is defined between "bra" ($_{l}\langle n|$) and
"ket" ($|m\rangle _{r}$) states
\begin{equation}
_{l}\langle n|m\rangle _{r}=\delta _{nm}.  \label{C-1}
\end{equation}

The eigenvalues of Hamiltonian are given by%
\begin{equation}
\widehat{H}|n\rangle _{r}=i\omega \hbar \left( n+\frac{1}{2}\right)
|n\rangle _{r},\quad \widehat{H}|n\rangle _{l}=-i\omega \hbar \left( n+\frac{%
1}{2}\right) |n\rangle _{l}.  \label{D}
\end{equation}

and the anti-normal order is%
\begin{equation}
_{r}\langle n|\widehat{H}=-i\omega \hbar \left( n+\frac{1}{2}\right)
_{r}\langle n|,\quad _{l}\langle n|\widehat{H}=i\omega \hbar \left( n+\frac{1%
}{2}\right) _{l}\langle n|.  \label{E}
\end{equation}

The expectation values are defined as%
\begin{equation*}
_{l}\langle n|\widehat{H}|n\rangle _{r}=i\omega \hbar \left( n+\frac{1}{2}%
\right) ,\quad _{r}\langle n|\widehat{H}|n\rangle _{l}=-i\omega \hbar \left(
n+\frac{1}{2}\right) ,
\end{equation*}%
which are unique. The expectation value on the "bra" or "ket" itself depends
on the normal/anti-normal orders.%
\begin{equation*}
_{r}\langle n|\overrightarrow{\overleftarrow{H}}|n\rangle _{r}=\pm i\omega
\hbar \left( n+\frac{1}{2}\right) \quad _{r}\langle n|n\rangle _{r},\quad
_{l}\langle n|\overrightarrow{\overleftarrow{H}}|n\rangle _{l}=\mp i\omega
\hbar \left( n+\frac{1}{2}\right) \quad _{l}\langle n|n\rangle _{l},
\end{equation*}%
where the probability densities $_{i}\langle n|n\rangle _{i}$ with $i=r,l$
are non-normalized. They are not the conserved quantities either because the
unstable nature of the eigenstates.

\subsection{ $SU(1,1)-$ generator representation}

The $SU(1,1)$ generators can be realized by the imaginary frequency
boson-operators such as%
\begin{eqnarray}
\widehat{S}_{z} &=&\frac{1}{2}(\widehat{a}_{+}\widehat{a}_{-}+\frac{1}{2}%
),\quad \widehat{S}_{z}^{\dag }=-\widehat{S}_{z}  \notag \\
\widehat{S}_{\pm } &=&\frac{1}{2}\widehat{a}_{\pm }^{2},\quad \widehat{S}%
_{\pm }^{\dag }=-\widehat{S}_{\pm },  \label{F}
\end{eqnarray}%
in which $\widehat{S}_{z}$ is anti-Hermitian different from the case of
normal oscillator \cite{LW2023}. The operators of $x,y$ components are
defined as usual%
\begin{eqnarray*}
\widehat{S}_{x} &=&\frac{1}{2}\left( \widehat{S}_{+}+\widehat{S}_{-}\right) ,
\\
\widehat{S}_{y} &=&\frac{1}{2i}\left( \widehat{S}_{+}-\widehat{S}_{-}\right)
.
\end{eqnarray*}%
The $x$ component is anti-Hermitian%
\begin{equation*}
\widehat{S}_{x}^{\dag }=-\widehat{S}_{x}
\end{equation*}%
while the $y$ component is Hermitian%
\begin{equation*}
\widehat{S}_{y}^{\dag }=\widehat{S}_{y}.
\end{equation*}%
The $SU(1,1)$ commutation relation is satisfied such that%
\begin{eqnarray*}
\left[ \widehat{S}_{x},\quad \widehat{S}_{y}\right]  &=&i\widehat{S}_{z}, \\
\left[ \widehat{S}_{z},\quad \widehat{S}_{\pm }\right]  &=&\pm \widehat{S}%
_{\pm }, \\
\left[ \widehat{S}_{+},\quad \widehat{S}_{-}\right]  &=&-2\widehat{S}_{z}.
\end{eqnarray*}%
The Hamiltonian is represented by the $SU(1,1)$ generator as%
\begin{equation}
\widehat{H}=2i\omega \hbar \widehat{S}_{z}.  \label{G}
\end{equation}

\subsection{\protect\bigskip Raising and lowering operators, recurrence
relation of eigenstates}

We present the recurrence relation of eigenstates of $\widehat{n}$ and $%
\widehat{n}^{\dag }$ with the lowering and raising operators. The operator $%
\widehat{a}_{-}$ acts on the "ket" states as a lowering operator since
\begin{equation*}
\widehat{n}\widehat{a}_{-}|n\rangle _{r}=\left( n-1\right) \widehat{a}%
_{-}|n\rangle _{r}
\end{equation*}%
according to the commutation relation of imaginary-frequency boson operator
Eq.(\ref{A-1}) Thus
\begin{equation}
\widehat{a}_{-}|n\rangle _{r}=c_{n-}^{r}|n-1\rangle _{r}.  \label{H}
\end{equation}%
$\widehat{a}_{+}$ is the raising operator of the "ket" states,
\begin{equation*}
\widehat{n}\widehat{a}_{+}|n-1\rangle _{r}=n\widehat{a}_{+}|n-1\rangle _{r}
\end{equation*}%
and%
\begin{equation}
\widehat{a}_{+}|n-1\rangle _{r}=c_{\left( n-1\right) +}^{r}|n\rangle _{r}.
\label{I}
\end{equation}%
Using Eqs.(\ref{H}, \ref{I}), we have
\begin{equation*}
\widehat{n}|n\rangle _{r}=n|n\rangle _{r}=c_{n-}^{r}c_{\left( n-1\right)
+}^{r}|n\rangle _{r}.
\end{equation*}%
then%
\begin{equation*}
c_{n-}^{r}c_{\left( n-1\right) +}^{r}=n,
\end{equation*}%
from the normalization condition Eq.(\ref{C-1}). Repeating the same
procedure on the states $_{-}|n-1\rangle _{r}$ , $_{-}|n-2\rangle _{r}$ and
so on, we obtain the recurrence relation

\begin{equation*}
c_{\left( n-1\right) -}^{r}c_{\left( n-2\right) +}^{r}=n-1
\end{equation*}%
\begin{equation*}
\cdot \cdot \cdot \cdot \cdot \cdot
\end{equation*}%
\begin{equation*}
c_{1-}^{r}c_{0+}^{r}=1
\end{equation*}%
The state $_{-}|0\rangle _{r}$ corresponds to the lowest initial energy and
thus the lowering operation must be terminated such that
\begin{equation}
\widehat{a}_{-}|0\rangle _{r}=0,\quad c_{0-}^{r}=0  \label{J}
\end{equation}%
We set the solution%
\begin{equation}
c_{n-}^{r}=c_{\left( n-1\right) +}^{r}=\sqrt{n},c_{\left( n-1\right)
-}^{r}=c_{\left( n-2\right) +}^{r}=\sqrt{n-1}...c_{1-}^{r}=c_{0+}^{r}=1.
\label{J-1}
\end{equation}%
Then the $n$-th eigenstate $|n\rangle _{r}$ can be generated from the ground
state $|0\rangle _{r}$ by the raising operator
\begin{equation}
|n\rangle _{r}=\frac{(\widehat{a}_{+})^{n}}{\sqrt{n!}}|0\rangle _{r}
\label{K}
\end{equation}%
For the "bra" states the number operator is $\widehat{n}^{\dag }$. It is
easy to find from Eqs.(\ref{C}, \ref{C-2}) that
\begin{equation*}
\widehat{n}^{\dag }a_{-}|n\rangle _{l}=\left( n+1\right) \widehat{a}%
_{-}|n\rangle _{l},
\end{equation*}%
and%
\begin{equation*}
\widehat{n}^{\dag }\widehat{a}_{+}|n\rangle _{l}=\left( n-1\right) \widehat{a%
}_{+}|n\rangle _{l}.
\end{equation*}%
$\widehat{a}_{-}$ becomes the raising operator while lowering operator is $%
\widehat{a}_{+}$ opposite to the case of "ket" state. We assume%
\begin{equation*}
\widehat{a}_{-}|n-1\rangle _{l}=c_{(n-1)-}^{l}|n\rangle _{l},
\end{equation*}%
and%
\begin{equation*}
\widehat{a}_{+}|n\rangle _{l}=c_{n+}^{l}|n-1\rangle _{l}.
\end{equation*}%
Then we have according to Eqs.(\ref{C}, \ref{C-2})
\begin{equation*}
\widehat{n}^{\dag }|n\rangle _{l}=n|n\rangle
_{l}=-c_{(n-1)-}^{l}c_{n+}^{l}|n\rangle _{l}.
\end{equation*}%
Following the same procedure as in the "ket" states we obtain%
\begin{equation*}
c_{n+}^{l}c_{\left( n-1\right) -}^{l}=-n
\end{equation*}%
\begin{equation*}
c_{(n-1)+}^{l}c_{\left( n-2\right) -}^{l}=-\left( n-1\right)
\end{equation*}%
\begin{equation*}
\cdot \cdot \cdot \cdot \cdot \cdot
\end{equation*}%
\begin{equation}
c_{1+}^{l}c_{0-}^{l}=-1  \label{L}
\end{equation}%
The lowering operation on the lowest energy state $|0\rangle _{l}$ must be
terminated%
\begin{equation}
\widehat{a}_{+}|0\rangle _{l}=0,\quad c_{0+}^{l}=0.  \label{L-1}
\end{equation}%
We assume the solutions of Eqs.(\ref{L})
\begin{equation}
c_{n+}^{l}=c_{\left( n-1\right) -}^{l}=-i\sqrt{n},\quad
c_{(n-1)+}^{l}=c_{\left( n-2\right) -}^{l}=-i\sqrt{n-1}%
...c_{1+}^{l}=c_{0-}^{l}=-i.  \label{L-3}
\end{equation}%
The $n$-th eigenstate is generated from the ground state by the raising
operator%
\begin{equation}
|n\rangle _{l}=\frac{\left( \widehat{a}_{-}\right) ^{n}}{(-i)^{n}\sqrt{n!}}%
|0\rangle _{l}  \label{L-2}
\end{equation}

\section{Wave functions and the normalization}

The generating function of the "ket" wave function is obtained from the
ground state equation Eq.(\ref{J})%
\begin{equation*}
\widehat{a}_{-}|0\rangle _{r}=0,
\end{equation*}%
which in coordinate representation is
\begin{equation*}
\sqrt{\frac{i}{2}}\int \langle x|\left( \widehat{x}+\widehat{p}\right)
|x^{\prime }\rangle \langle x^{\prime }|0\rangle _{r}dx^{\prime }=0
\end{equation*}%
namely%
\begin{equation*}
\left( x-i\frac{d}{dx}\right) \psi _{0}^{r}=0.
\end{equation*}%
The ground state wave function, called the generating function, is%
\begin{equation}
\psi _{0}^{r}=\frac{1}{\sqrt{N_{r}}}e^{-i\frac{x^{2}}{2}},  \label{M}
\end{equation}%
where $N_{r}$ is the normalization constant to be determined. The $n$-th
excited wave function can be generated by the raising operator from Eq.(\ref%
{K})
\begin{equation}
\psi _{n}^{r}=\frac{1}{\sqrt{N_{r}}\sqrt{n!}}\left( \sqrt{\frac{i}{2}}%
\right) ^{n}\left( x-i\frac{d}{dx}\right) ^{n}e^{-i\frac{x^{2}}{2}}
\label{S}
\end{equation}

The generating function of "bra" states is derived from the equation Eq.(\ref%
{L-1})%
\begin{equation*}
\widehat{a}_{+}|0\rangle _{l}=0,
\end{equation*}%
which becomes in the coordinate representation%
\begin{equation*}
\left( x+i\frac{d}{dx}\right) \psi _{0}^{l}=0.
\end{equation*}%
Then%
\begin{equation*}
\psi _{0}^{l}=\frac{1}{\sqrt{N_{l}}}e^{i\frac{x^{2}}{2}}.
\end{equation*}%
The $n$-th "bra" eigenstate wave function is written as%
\begin{equation}
\psi _{n}^{l}=\frac{1}{(-i)^{n}\sqrt{n!}\sqrt{N_{l}}}\left( \sqrt{\frac{i}{2}%
}\right) ^{n}\left( x+i\frac{d}{dx}\right) ^{n}e^{i\frac{x^{2}}{2}}
\label{X}
\end{equation}%
The normalization condition is obtained with the help of imaginary
integration measure used in the Feynman path integral \cite{FMP,LW2023}
\begin{equation*}
\int_{l}\langle 0|x\rangle \langle x|0\rangle _{r}dx=\frac{1}{\sqrt{%
N_{l}^{\ast }N_{r}}}\int e^{-ix^{2}}dx=\frac{1}{\sqrt{N_{l}^{\ast }N_{r}}}%
\sqrt{\frac{\pi }{i}}=1,
\end{equation*}%
which gives rise to the normalization constant
\begin{equation*}
N_{r}=\sqrt{\frac{\pi }{i}},\quad N_{l}=\sqrt{\frac{\pi }{-i}}
\end{equation*}%
The normalized wave functions are given by%
\begin{equation*}
\psi _{0}^{r}=\left( \frac{i}{\pi }\right) ^{\frac{1}{4}}e^{-i\frac{x^{2}}{2}%
},\psi _{0}^{l}=\left( \frac{-i}{\pi }\right) ^{\frac{1}{4}}e^{i\frac{x^{2}}{%
2}}
\end{equation*}%
\begin{eqnarray}
\psi _{n}^{r} &=&\frac{1}{\sqrt{n!}}\left( \sqrt{\frac{i}{2}}\right)
^{n}\left( \frac{i}{\pi }\right) ^{\frac{1}{4}}\left( x-i\frac{d}{dx}\right)
^{n}e^{-i\frac{x^{2}}{2}}  \notag \\
\psi _{n}^{l} &=&\frac{1}{(-i)^{n}\sqrt{n!}}\left( \sqrt{\frac{i}{2}}\right)
^{n}\left( \frac{-i}{\pi }\right) ^{\frac{1}{4}}\left( x+i\frac{d}{dx}%
\right) ^{n}e^{i\frac{x^{2}}{2}}  \label{Y}
\end{eqnarray}%
The generating function is normalized with the imaginary integration measure%
\begin{equation*}
\int (\psi _{0}^{l})^{\ast }\psi _{0}^{r}dx=\frac{1}{\sqrt{\pi }}\int e^{-(%
\sqrt{i}x)^{2}}d(\sqrt{i}x)=1,
\end{equation*}%
exactly the same as in the Feynman path integral \cite{FMP,LW2023}. While
both the probabilities of "bra" and "ket" states are constant
\begin{equation*}
(\psi _{0}^{i})^{\ast }\psi _{0}^{i}=\frac{1}{\sqrt{\pi }},
\end{equation*}%
for $i=r,l$. The states are non-normalizable since the space-coordinate
integration leads to infinity%
\begin{equation*}
\int (\psi _{0}^{i})^{\ast }\psi _{0}^{i}dx\rightarrow \infty .
\end{equation*}

\section{Coherent state of imaginary-frequency boson operator}

The coherent state of imaginary-frequency boson operator can be defined as
the eigenstate of the lowering operators $\widehat{a}_{-}$ , $\widehat{a}%
_{+} $ respectively for the "ket" and "bra" sets of eigenstates.

\subsection{Coherent state of "ket" set}

The coherent state of "ket" set defined as the eigenstate of lowering
operator%
\begin{equation}
\widehat{a}_{-}|\alpha \rangle _{r}=\alpha |\alpha \rangle _{r},  \label{2}
\end{equation}%
can be written as the expansion of eigenstates of number operator $\widehat{n%
}=\widehat{a}_{+}\widehat{a}_{-}$ ,%
\begin{equation*}
|\alpha \rangle _{r}=\sum_{n}c_{n}^{r}|n\rangle _{r}.
\end{equation*}%
The definition Eq.(\ref{2}) leads to recurrence equation
\begin{equation}
\widehat{a}_{-}|\alpha \rangle _{r}=\sum_{n}c_{n}^{r}c_{n-1}^{r}|n-1\rangle
_{r}=\alpha \sum_{n}c_{n}^{r}|n\rangle _{r}.  \label{1}
\end{equation}%
Then applying the raising operator on Eq.(\ref{1}) we have%
\begin{equation}
\sum_{n}c_{n}^{r}\widehat{a}_{+}\widehat{a}_{-}|n\rangle
_{r}=\sum_{n}c_{n}^{r}n|n\rangle _{r}=\alpha
\sum_{n}c_{n}^{r}c_{n+}^{r}|n+1\rangle _{r}.  \label{3}
\end{equation}%
Taking the inner product with the state $_{l}\langle n|$ we obtain
\begin{equation}
c_{n}^{r}n=\alpha c_{n-1}^{r}c_{(n-1)+}^{r}.  \label{4}
\end{equation}%
Since $c_{(n-1)+}^{r}=\sqrt{n}$ from Eq.(\ref{J-1}), the recurrence relation
of coefficients is%
\begin{equation*}
c_{n}^{r}=\frac{\alpha }{\sqrt{n}}c_{n-1}^{r},
\end{equation*}%
\begin{equation*}
c_{n-1}^{r}=\frac{\alpha }{\sqrt{n-1}}c_{n-2}^{r}
\end{equation*}

\begin{equation*}
\cdot \cdot \cdot \cdot \cdot \cdot
\end{equation*}%
\begin{equation}
c_{1}^{r}=\alpha c_{0}^{r}  \label{5}
\end{equation}%
The coherent state becomes%
\begin{equation}
|\alpha \rangle _{r}=\sum_{n}\frac{\alpha ^{n}}{\sqrt{n!}}c_{0}^{r}|n\rangle
_{r}  \label{6}
\end{equation}%
with the parameter $c_{0}^{r}$ to be determined from the normalization
condition.

\subsection{Coherent state of "bra" set}

The coherent state for "bra" set is defined as the eigenstate of lowering
operator such as%
\begin{equation}
\widehat{a}_{+}|\alpha \rangle _{l}=\alpha |\alpha \rangle _{l},\quad
|\alpha \rangle _{l}=\sum_{n}c_{n}^{l}|n\rangle _{l},  \label{6-1}
\end{equation}%
and%
\begin{equation}
\sum c_{n}^{l}\widehat{a}_{+}|n\rangle _{l}=\alpha \sum c_{n}^{l}|n\rangle
_{l}.  \label{7}
\end{equation}%
Applying the raising operator $-\widehat{a}_{-}$on \ Eq.(\ref{7}) yields%
\begin{equation*}
\sum c_{n}^{l}\left( -\widehat{a}_{-}\widehat{a}_{+}\right) |n\rangle
_{l}=\sum c_{n}^{l}\widehat{n}^{\dag }|n\rangle _{l}=-\alpha \sum c_{n}^{l}%
\widehat{a}_{-}|n\rangle _{l},
\end{equation*}%
which becomes%
\begin{eqnarray}
\sum c_{n}^{l}n|n\rangle _{l} &=&-\alpha \sum c_{n}^{l}c_{n-}^{l}|n+1\rangle
_{l}=-\alpha \sum c_{n-1}^{l}c_{(n-1)-}^{l}|n\rangle _{l}  \notag \\
&=&\alpha \sum c_{n-1}^{l}i\sqrt{n}|n\rangle _{l},  \label{8}
\end{eqnarray}%
noticing the value $c_{(n-1)-}^{l}=-i\sqrt{n}$ in Eq.(\ref{L-3}). The
recurrence relation of coefficients is%
\begin{eqnarray*}
c_{n}^{l} &=&\frac{i\alpha }{\sqrt{n}}c_{\left( n-1\right) }^{l}, \\
c_{n-1}^{l} &=&\frac{i\alpha }{\sqrt{n-1}}c_{\left( n-2\right) }^{l},
\end{eqnarray*}%
\begin{equation*}
\cdot \cdot \cdot \cdot \cdot \cdot
\end{equation*}

\begin{equation}
c_{1}^{l}=i\alpha c_{0}^{l}  \label{9}
\end{equation}

The coherent state of "bra" set is%
\begin{equation}
|\alpha \rangle _{l}=\sum_{n}\frac{\left( i\alpha \right) ^{n}}{\sqrt{n!}}%
c_{0}^{l}|n\rangle _{l}.  \label{10}
\end{equation}%
The normalization condition between the "bra" and "ket" coherent states is
seen to be%
\begin{eqnarray*}
_{l}\langle \alpha |\alpha \rangle _{r} &=&\sum_{n}\frac{(-i)^{n}|\alpha
|^{2n}}{n!}\left( c_{0}^{l}\right) ^{\ast }c_{0}^{r} \\
&=&e^{-i|\alpha |^{2}}\left( c_{0}^{l}\right) ^{\ast }c_{0}^{r}=1
\end{eqnarray*}%
The normalization constant is%
\begin{equation*}
c_{0}^{r}=e^{i\frac{|\alpha |^{2}}{2}},\quad c_{0}^{l}=e^{-i\frac{|\alpha
|^{2}}{2}}.
\end{equation*}%
The normalized coherent states are given by%
\begin{equation}
|\alpha \rangle _{r}=e^{i\frac{|\alpha |^{2}}{2}}\sum_{n}\frac{\alpha ^{n}}{%
\sqrt{n!}}|n\rangle _{r}  \label{11}
\end{equation}%
and%
\begin{equation}
|\alpha \rangle _{l}=e^{-i\frac{|\alpha |^{2}}{2}}\sum_{n}\frac{\left(
-i\alpha \right) ^{n}}{\sqrt{n!}}|n\rangle _{l}.  \label{12}
\end{equation}

\subsection{Minimum uncertainty}

The coherent state can be also defined by the minimum uncertainty. The
position and momentum operators are represented by imaginary frequency boson
operators respectively as%
\begin{equation}
\widehat{x}=\frac{1}{\sqrt{2i}}\left( \widehat{a}_{-}+\widehat{a}_{+}\right)
,\quad \widehat{p}=\frac{1}{\sqrt{2i}}\left( \widehat{a}_{-}-\widehat{a}%
_{+}\right) .  \label{13}
\end{equation}%
According to the definition of coherent states
\begin{equation*}
\widehat{a}_{-}|\alpha \rangle _{r}=\alpha |\alpha \rangle _{r},\quad
_{r}\langle \alpha |\widehat{a}_{-}^{\dag }=\quad _{r}\langle \alpha |\alpha
^{\ast },\quad \widehat{a}_{-}^{\dag }=i\widehat{a}_{-},\quad
\end{equation*}%
\begin{equation*}
\widehat{a}_{+}|\alpha \rangle _{l}=\alpha |\alpha \rangle _{l},\quad
_{l}\langle \alpha |\widehat{a}_{+}^{\dag }=\quad _{l}\langle \alpha |\alpha
^{\ast },\quad \widehat{a}_{+}^{\dag }=i\widehat{a}_{+},\quad
\end{equation*}%
the average of position operator on the coherent state is evaluated as%
\begin{equation}
\langle \widehat{x}\rangle =\frac{1}{\sqrt{2i}}\quad _{l}\langle \alpha
|\left( \widehat{a}_{-}+\widehat{a}_{+}\right) |\alpha \rangle _{r}=\frac{1}{%
\sqrt{2i}}\left( \alpha -i\alpha ^{\ast }\right) ,  \label{14}
\end{equation}%
and the square of average is%
\begin{equation}
\langle \widehat{x}\rangle ^{2}=\frac{1}{2i}\left( \alpha ^{2}-2i|\alpha
|^{2}-\left( \alpha ^{\ast }\right) ^{2}\right)  \label{15}
\end{equation}%
While the square of position operator reads%
\begin{equation*}
\widehat{x}^{2}=\frac{1}{2i}\left( \widehat{a}_{+}^{2}+2\widehat{a}_{+}%
\widehat{a}_{-}+1+\widehat{a}_{-}^{2}\right)
\end{equation*}%
with the average given by%
\begin{equation}
\langle \widehat{x}^{2}\rangle =\frac{1}{2i}\left( -\left( \alpha ^{\ast
}\right) ^{2}-i2\alpha ^{\ast }\alpha +1+\alpha _{-}^{2}\right)  \label{16}
\end{equation}%
The deviation is%
\begin{equation*}
\langle \widehat{x}^{2}\rangle -\langle \widehat{x}\rangle ^{2}=-\frac{i}{2}.
\end{equation*}%
The expectation value of momentum operator is%
\begin{equation*}
\langle \widehat{p}\rangle =\frac{1}{\sqrt{2i}}\quad _{l}\langle \alpha |%
\widehat{a}_{-}-\widehat{a}_{+}|\alpha \rangle _{r}\quad =\frac{1}{\sqrt{2i}}%
\left( \alpha +i\alpha ^{\ast }\right)
\end{equation*}%
with square given by%
\begin{equation}
\langle \widehat{p}\rangle ^{2}=\frac{1}{2i}\left( \alpha _{-}^{2}-\left(
\alpha ^{\ast }\right) ^{2}+i2\alpha ^{\ast }\alpha \right) .  \label{17}
\end{equation}%
The expectation value of momentum square is%
\begin{equation*}
\langle \widehat{p}^{2}\rangle =\frac{1}{2i}\left( \alpha _{-}^{2}-\left(
\alpha ^{\ast }\right) ^{2}+i2\alpha ^{\ast }\alpha -1\right) .
\end{equation*}%
The\bigskip\ momentum deviation is%
\begin{equation*}
\langle \widehat{p}^{2}\rangle -\langle \widehat{p}\rangle ^{2}=\frac{i}{2}
\end{equation*}%
The minimum uncertainty is satisfied,%
\begin{equation}
\Delta x\Delta p=\frac{1}{2},  \label{18}
\end{equation}%
where $\Delta x=\sqrt{\langle \widehat{x}^{2}\rangle -\langle \widehat{x}%
\rangle ^{2}},$ and $\Delta p=\sqrt{\langle \widehat{p}^{2}\rangle -\langle
\widehat{p}\rangle ^{2}}$.

\section{Decay of eigenstates and dynamic quantum-classical correspondence}

The eigenstate with imaginary eigenvalue is unstable. The imaginary
eigenvalue characterizes the decay rate. The special solutions of
time-dependent Schr$\ddot{o}$dinger equations Eq.(\ref{s}) are given by%
\begin{equation}
|\psi _{n}\left( t\right) \rangle _{r}=|n\rangle _{r}e^{\left( n+\frac{1}{2}%
\right) \omega t},\quad |\psi _{n}\left( t\right) \rangle _{l}=|n\rangle
_{l}e^{-\left( n+\frac{1}{2}\right) \omega t}  \label{19}
\end{equation}%
respectively for the "ket" and "bra" states. The probability density is no
longer a conserved quantity. It increases or decreases with time indicating
the nature of imaginary eigenvalue. The non-Hermitian density operator
\begin{equation*}
\widehat{\rho }_{r,l}=|\psi \rangle _{rl}\langle \psi |
\end{equation*}%
for the "bra" and "ket" is an invariant%
\begin{equation}
i\hbar \frac{d}{dt}\widehat{\rho }_{r,l}=i\hbar \frac{\partial }{\partial t}%
\widehat{\rho }_{r,l}+\left[ \widehat{\rho }_{r,l},\quad \widehat{H}\right]
=0  \label{20}
\end{equation}%
which is consistent with the Schr$\ddot{o}$dinger equation Eq.(\ref{19})
\cite{GuResult,GuAnn,LiuResult,LW2023}. It is well known that the average of
Heisenberg equation in the coherent state coincides precisely with the
classical orbit for the normal oscillator \cite{LW2023}. We now investigate
the quantum dynamics of inverted potential well in the imaginary-frequency
boson coherent-states defined in Eqs.(\ref{2}, \ref{6-1}). To this end we
begin with Heisenberg equation%
\begin{equation*}
\frac{d\widehat{x}}{dt}=\frac{1}{i\hbar }\left[ \widehat{x},\quad \widehat{H}%
\right] =\omega \widehat{p},\quad \frac{d\widehat{p}}{dt}=\omega \widehat{x}%
,\quad
\end{equation*}%
and thus%
\begin{equation}
\frac{d^{2}\widehat{x}}{dt^{2}}=\omega ^{2}\widehat{x}.  \label{21}
\end{equation}%
The expectation value of Heisenberg Eq.(\ref{21}) in the coherent state is%
\begin{equation}
\overset{..}{\alpha }-i(\overset{..}{\alpha })^{\ast }=\omega ^{2}\left(
\alpha -i\alpha ^{\ast }\right)  \label{22}
\end{equation}%
respectively for the average $_{l}\langle \alpha |\widehat{x}|\alpha \rangle
_{r}$ and%
\begin{equation}
\overset{..}{\alpha }+i(\overset{..}{\alpha })^{\ast }=\omega ^{2}\left(
\alpha +i\alpha ^{\ast }\right)  \label{23}
\end{equation}%
for the average $_{r}\langle \alpha |\widehat{x}|\alpha \rangle _{l}$. Eqs.(%
\ref{22}, \ref{23}) lead to the differential equation
\begin{equation}
\overset{..}{\alpha }=\omega ^{2}\alpha ,  \label{24}
\end{equation}%
the solution of which is exactly the same of classical orbit Eq.(\ref{c-1})
\begin{equation*}
\overline{x}\left( t\right) =\alpha \left( t\right) =\pm \frac{v}{\omega }%
\sinh \omega t.
\end{equation*}%
The quantum-classical correspondence exists also in the system with
imaginary eigenvalues.

\section{Conclusion}

For the system of a particle in an inverted potential well the Hamiltonian
is Hermitian, however, the potential is unbounded below. The eigenvalues are
pure imaginary since the eigenstates are unstable. The decay of metastable
state was studied long ago with instanton method, in which the imaginary
part of complex eigenvalue characterizes the decay rate \cite%
{LM92,LM94,Liang}. The Hermitian Hamiltonian with pure imaginary eigenvalues
has not yet been formulated self-consistently under the basic commutation
relation of canonical-variable operators. The inverted potential well
problem is solved in the present paper by means of the algebraic method with
the imaginary-frequency boson operators. Dual sets of eigenstates are
necessary corresponding to the complex conjugate eigenvalues. The
orthonormal condition is valid between the dual sets of eigenstates. The
boson number operator is non-Hermitian while the integer-number eigenvalues
are, of course, real. The probability density of wave function is spatially
non-localized with the normalization defined under the imaginary measure of
integration, the same as the Feynman path integrals \cite{FMP,LW2023}. The
boson coherent states derived as the eigenstates of lowering operators
satisfy explicitly the minimum uncertainty of momentum and position
operators. The expectation value of Heisenberg equation in the coherent
states gives rise to the same classical orbit indicating exact
quantum-classical correspondence.

\end{document}